\documentstyle[preprint,revtex]{aps}
\begin{document}
\draft
\input{psfig}
\def\rslide#1#2{\centerline{\psfig{figure=#1,height=#2,bbllx=-774bp,bblly=68bp,
bburx=-18bp,bbury=514bp,clip=}} }
\preprint{RUB-TPII-21/92 \\}
\preprint{June 1992}
\begin{title}
ON THE NUCLEON DISTRIBUTION AMPLITUDE: \\
           THE HETEROTIC SOLUTION
\end{title}
\author{N. G. Stefanis and M. Bergmann}
\begin{instit}
Institut f\"ur Theoretische Physik II  \\
Ruhr-Universit\"at Bochum  \\
D-4630 Bochum, Germany
\end{instit}

\begin{abstract}
We present a new nucleon distribution amplitude which
amalgamates features of the Chernyak-Ogloblin-Zhitnitsky model with
those of the Gari-Stefanis model. This "heterotic" solution provides
the possibility to have asymptotically a small ratio
\hbox{$\vert G_{M}^{n}\vert/G_{M}^{p}\le 0.1$},
while fulfilling most of the sum-rule requirements up to the third
order. Using this nucleon distribution amplitude we calculate the
electromagnetic and weak nucleon form factors, the transition form
factor $\gamma p \Delta^{+}$ and the decay widths of the charmonium
levels $^3S_{1}$, $^3P_{1}$, and $^3P_{2}$ into $p\bar p$. The
agreement with the available data is remarkable in all cases.
\end{abstract}
\pacs{PACS numbers: 12.38.Bx, 12.38.Lg, 13.25.+m, 13.40.Fn \\ \\
      Internet: michaelb@photon.tp2.ruhr-uni-bochum.de \\
\phantom{Internet:\ }nicos@hadron.tp2.ruhr-uni-bochum.de \\
Bitnet: KPH509@DJUKFA11  }
\narrowtext
In recent years a number of authors~\cite{CZ,GS,KS,COZ1}
have proposed various models for nucleon distribution
amplitudes (NDA), based on light-cone
perturbative QCD~\cite{LB,AKC} in
conjunction with QCD sum rules (SR)~\cite{CZ,KS,COZ1}.
With varying degrees
of conviction these models have been used in a series of analyses to
calculate the electromagnetic~\cite{CZ,GS,COZ1,JSL,COZ2,S}
and the weak~\cite{CP1,CP2} nucleon form factors, the transition
form factor $\gamma p\Delta^{+}$~\cite{C,CGS,WC,STO,COZ2},
the cross section for proton Compton scattering at large momentum
transfer $Q^2$~\cite{KN} and the exclusive $p\bar p$ decays
of heavy quarkonia~\cite{CZ,COZ2,A}. They may also be useful
in determining the pion form factor at large spacelike $Q^2$ via
pion electroproduction~\cite{CM}.

Although these models incorporate essential ingredients for a unified
description of perturbative and nonperturbative aspects of the
subnucleon structure, important questions still remain unanswered.
One question focuses on the value of $G_{M}^{n}/{G_{M}^{p}}$
and the possibility that the electron-neutron differential cross
section $\sigma_{n}$ is dominated by $G_{E}^{n}$, while $G_{M}^{n}$ is
asymptotically small or equivalently that $\vert F_{1}^{n}\vert\ll
\vert F_{2}^{n}\vert$ at all $Q^2$ values
\cite{KK}.
The Gari-Stefanis (GS) amplitude~\cite{GS} was constructed to
account for this behavior of the form factors and gives good agreement
with the latest high-$Q^2$ SLAC data~\cite{SLAC} at the expense
that the moments $(002)$ and $(101)$ cannot match the SR
requirements~\cite{CZ} of Chernyak and Zhitnitsky (CZ) in the allowed
saturation range~\cite{S}. A second issue concerns whether higher
terms in the Appell polynomial decomposition of the
NDA~\cite{LB} are significant, a point that
was raised in~\cite{SCH}. A systematic investigation
of this matter will be conducted elsewhere. In the present work
we resort to second-order Appell polynomials.

In~\cite{COZ1} Chernyak, Ogloblin, and Zhitnitsky (COZ) have
recalculated the SR for the first- and second-order
moments of the NDA and have derived
for the first time SR for the third-order moments.
Their new set of SR comprises $18$ terms with restricted
margins which comply with the results of the King-Sachrajda (KS)
computation~\cite{KS} but disagree with those obtained on the
lattice for the lowest two moments~\cite{RSS}. They have also
proposed a new NDA which satisfies
all, but 6 SR, whereas the CZ amplitude and the GS
amplitude violate, respectively, 13 and 14 of the new SR.
The KS amplitude provides almost the same quality as the COZ amplitude
with only 7 SR being broken.

In~\cite{COZ2} the same authors pointed out that the GS model
leads to a prediction for the $^3S_{1} \to p\bar p$ decay width which
is about $50$ times smaller than the experimental value. This, in
connection with the strong violation of the SR for the moment
(002)
led Chernyak, Ogloblin, and Zhitnitsky to the conclusion that the
GS model is unacceptable. On the other hand, they
stressed~\cite{COZ1,COZ2} that "they have not succeeded in
finding the model distribution amplitude which contains not
higher than second-order Appell polynomials and gives
$\vert F_{1}^{n}\vert/F_{1}^{p} <0.4$ while fulfilling the
SR."

It is the purpose of this note to present a unique NDA
which makes it possible to resolve all
these problems. This novel solution turns out to be something of
a hybrid between the COZ amplitude and the GS amplitude;
thus the "heterotic" NDA. Specifically,
we find $\vert G_{M}^{n}\vert/G_{M}^{p}\le 0.1$ and a good
agreement between the calculated proton form factor
and the high-$Q^2$ data~\cite{SLAC}, whereas the fit
to the SR has almost the same accuracy as the
original COZ model. In addition, the heterotic distribution
amplitude leads to predictions for the exclusive decays of
the charmonium levels $^3S_{1}$, $^3P_{1}$, and $^3P_{2}$
into $p\bar p$, which are in remarkable agreement with the existing
experimental data.

The theoretical
basis for the description of hadronic exclusive processes
within QCD is provided by the factorization property (see, e.g.,
\cite{LB}), meaning that all soft (nonperturbative) effects
can be absorbed into quark distribution amplitudes for the hadrons
in the initial and final states, while the hard-scattering
subprocesses can be calculated via perturbative QCD.
The leading-order definition of the NDA,
modulo logarithmic corrections due to renormalization,
is
\begin{equation}
   \Phi(x_{i},Q^{2}) = \int_{}^{Q^{2}}[d^{2}k_{\bot}]\,
   \psi(x_{i},{\bf k}_{\bot}^{(i)})\;,
\end{equation}
where the measure is
\hbox{$[d^{2}k_{\bot}]=16\pi^{3}\delta^{(2)}(\sum_{i=1}^{3}\,
{\bf k}_{\bot}^{(i)})\prod_{i=1}^{3}\,\lbrack{{d^{2}k_{\bot}^{(i)}}
\over{16\pi^{3}}}\rbrack$} and $\psi(x_{i},{\bf k}_{\bot}^{(i)})$
is the lowest-twist Fock-space projection amplitude for finding
three valence quarks inside the nucleon, each carrying a
fraction $x_{i}={k_{i}^{+}}/{p^{+}}$ (with $p^{\pm}=p^{0}\pm p^{3}$)
of the nucleon's longitudinal momentum $p^{+}$
and having relative transverse momentum ${\bf k}_{\bot}^{(i)}$.
Although $\psi(x_{i},{\bf k}_{\bot}^{(i)})$, and hence
$\Phi(x_{i},Q^{2})$ cannot be calculated within perturbative
QCD, a $Q^{2}$ evolution equation can be derived from QCD
perturbation theory~\cite{LB}. Any solution of this equation
can be expressed in the form
\mediumtext
\begin{equation}
  \Phi(x_{i},Q^{2}) = \Phi_{as}(x_{i})\sum_{n=0}^{\infty}B_{n}(\mu^{2})
  \Biggl({{\alpha_{s}(Q^{2})}\over{\alpha_{s}(\mu^{2})}
}\Biggr)^{\gamma_{n}}
  \tilde \Phi_{n}(x_{i})\;,
\end{equation}
\narrowtext
where \hbox{$\Phi_{as}(x_{i})=120x_{1}x_{2}x_{3}$} and
$\{\tilde \Phi_{n}(x_{i})\}$ are the eigenfunctions of the
interaction kernel of the evolution equation, represented
in terms of Appell polynomials (cf. Ref.~\cite{LB}).
The corresponding eigenvalues $\gamma_{n}$ equal
the anomalous dimensions
of the lowest-twist three-quark operators with the appropriate
baryonic quantum numbers~\cite{P}.

The nonperturbative input enters Eq.(2) through the coefficients
$B_{n}(\mu^{2})$ which represent matrix elements of appropriate
three-quark operators (in the light-cone gauge $A^{+}=0$)
interpolating between the proton and the vacuum:
\mediumtext
\begin{equation}
  <\Omega\vert O_{\gamma}^{(n_{1}n_{2}n_{3})}(0)\vert P(p)> =
  f_{N}(z\cdot p)^{\,{n_{1}+n_{2}+n_{3}+1}}N_{\gamma}\,
  O^{\,(n_{1}n_{2}n_{3})}.
\end{equation}
\narrowtext
Here $z$ is a lightlike vector with $z^{2}=0$, $N_{\gamma}$ is the
proton spinor, and $f_{N}$ denotes the "proton decay constant".
SR calculations~\cite{CZ,KS,COZ1} make use of correlators
between two of the $O_{\,\gamma}^{(n_{1}n_{2}n_{3})}$:
\mediumtext
\begin{eqnarray}
I^{\,(n_{1}n_{2}n_{3},m)}(q,z) & = & i\int_{}^{}
  d^{4}x \, e^{iq\cdot x}
  <\Omega\vert T\bigl (O_{\gamma}^{\,(n_{1}n_{2}n_{3})}(0)
  \hat O_{\gamma\prime}^{\,(m)}(x)\bigr )\vert\Omega>(z\cdot \gamma)_
  {\gamma \gamma\prime} \nonumber\\
   & & \\
   &  = &
  (z\cdot q)^{\,{n_{1}+n_{2}+n_{3}+m+3}}I^{\,(n_{1}n_{2}n_{3},m)}(q^{2})\;.
  \nonumber
\end{eqnarray}
\narrowtext
To determine the moments of the NDA~\cite{CZ},
\begin{equation}
  \Phi_{N}^{\,(n_{1}n_{2}n_{3})} = \int_{0}^{1}[dx]\,
  x_{1}^{n_{1}}\,x_{2}^{n_{2}}\,x_{3}^{n_{3}}\,\Phi_{N}(x_{i})\;,
\end{equation}
(\hbox{$[dx]=dx_{1}dx_{2}dx_{3}\delta(1-x_{1}-x_{2}-x_{3})$})
a short-distance operator product expansion is performed at some
spacelike momentum $\mu^{2}$ where quark-hadron duality is valid.
By virtue of the orthogonality of the eigenfunctions $\tilde\Phi_{n}$,
the coefficients $B_{n}(\mu^{2})$ can be determined by inverting Eq. (5)
upon imposing the SR constraints. The moments
$\Phi_{N}^{(n_{1}n_{2}n_{3})}$ in terms of the coefficients $B_{n}$
for $n=0,1,\ldots 5$ have been given in~\cite{S}. Those for $n=6\ldots
9$ are calculated in~\cite{BS}.

Let us now outline our results. Treating the SR defined in (4)
for $n_{1}+n_{2}+n_{3}\le 3$ and $m=1$ within the range estimated
by COZ~\cite{COZ1}, the coefficients $B_{n}$ for the heterotic
solution are: $B_{0}=1$, $B_{1}=3.4437$, $B_{2}=1.5710$,
$B_{3}=4.5937$, $B_{4}=29.3125$, and $B_{5}=-0.1250$; the value
of $B_{0}$ is due to the normalization of $\Phi_{N}$, i.e.,
$\int_{0}^{1}[dx]\,\Phi_{N}(x_{i})=1$. The explicit form of
$\Phi_{N}^{het}$ is~\cite{S}:
\mediumtext
\begin{eqnarray}
 \Phi_{N}^{het}(x_{i}) &= \Phi_{as}(x_{i}) \bigl
 \{&  -19.773 +32.756(x_{1}-x_{3})+26.569x_{2} \nonumber\\
 & & \\
 & & +16.625x_{1}x_{3}
  -2.916x_{1}^{2}+75.25x_{3}^{2} \bigr \}\;.\nonumber
\end{eqnarray}
\narrowtext
The moments and the corresponding SR constraints are
shown in Tab.~\ref{Tab1}. There is a good overall consistency, with
only $7$ SR being broken. Given the fact that we have taken
into account only the first $6$ Appell polynomials to represent
$\Phi_{N}$, the deviations are tolerable.

The results for the magnetic form factor $G_{M}^{N}$ have
been obtained using analytical expressions given in~\cite{S}
and are plotted in Fig.~\ref{Figr2} for the proton
and Fig.~\ref{Figr3} for the neutron.
The data are from~\cite{SLAC,DATN}. The form-factor evolution
with $Q^{2}$ is due to the leading-order parametrization of the
effective coupling constant $\alpha_{s}(Q^{2})$. The evolution
of the coefficients $B_{n}$ is a minor effect and has been neglected.
Note that an average value
$\bar \alpha_{s}(Q^{2})=[\alpha_{s}(Q^{2}\times 0.427)
 \alpha_{s}(Q^{2}\times 0.178)]^{1/2}$
has been used to account for the different virtualities
of the involved propagators. Here and below two options
are shown corresponding to two typical values of the scale parameter
$\Lambda_{QCD}$,
while the proton decay constant is taken to be $\vert f_{N}\vert =
(5.0\pm0.3)\times 10^{-3}GeV^{2}$, as suggested by QCD-SR
\cite{CZ,KS,COZ1}.

Using $\Phi_{N}^{het}$ we have calculated the
electromagnetic $N-\Delta^{+}$ transition form factor
$G_{M}^{*}(Q^{2})$ for two~\cite{CP3,FZOZ} recently proposed
$\Delta^{+}$ distribution amplitudes (Fig.~\ref{Figr4}) following
\cite{C}. The solid and dashed curves refer to the model
of Carlson and Poor (CP)~\cite{CP3} for
$\Lambda_{QCD}=100 MeV$ and $\Lambda_{QCD} = 180 MeV$,
respectively. The dotted and dash-dotted curves are
their counterparts for the model of Farrar {\it et al.} (FZOZ)
\cite{FZOZ}. In all cases the CP value
$\vert f_{\Delta}\vert = 11.5\times 10^{-3} GeV^{2}$
has been used, which is within the spread of the FZOZ estimate.
The data are compiled in~\cite{DATD}. The $Q^{2}$ evolution of
$G_{M}^{*}(Q^{2})$ is governed by an effective coupling
constant, which we take to be the average of the two coupling
constants $\bar \alpha_{s}^{(N)}(Q^{2})$ and
$\bar \alpha_{s}^{(\Delta)}(Q^{2})$ with arguments weighted
by the virtualities appropriate to each model:
$\bar \alpha_{s}^{(CP)}(Q^{2})=[\alpha_{s}(Q^{2}\times 0.3773)
 \alpha_{s}(Q^{2}\times 0.1488)]^{1/2}$ and
$\bar \alpha_{s}^{(FZOZ)}(Q^{2})=[\alpha_{s}(Q^{2}\times 0.4643)
 \alpha_{s}(Q^{2}\times 0.1015)]^{1/2}$.
If we take these predictions at face value, then the available
data seems to favor the CP model (cf.~\cite{BS92}).
On the other hand, there is as yet
no definite experimental evidence whether the $Q^{4}G_{M}^{*}$
curve levels off or descends rapidly to zero, as predicted by COZ-type
NDA~\cite{CGS}.
A priori we have no reason to favor one option over the other
(and in fact the recent analysis by Stoler~\cite{STO}
of the unpublished data of the SLAC experiment E133 points to
the second possibility). Further exclusive experiments to measure
$G_{M}^{*}$ at as high $Q^{2}$ as possible are crucial.

The calculation for the nucleon axial form factor $g_{A}(Q^{2})$
according to~\cite{CP1} yields at $Q^{2}\approx 10 GeV^{2}$,
$Q^{4}g_{A}(Q^{2})=0.90 GeV^{4}$ for $\Lambda_{QCD}=100 MeV$,
and $Q^{4}g_{A}(Q^{2})=1.44 GeV^{4}$ for $\Lambda_{QCD} = 180 MeV$.
These results compare well with the value
$Q^{4}g_{A}(Q^{2})\approx 1.5 GeV^{4}$ extrapolated from the data
\cite{DATA}. Also the ratio ${g_{A}(Q^{2})}/{G_{M}^{p}(Q^{2})}
\approx 1.19$, in the region where the calculations can be trusted,
is consistent with the (extrapolated) experimental value
${g_{A}(Q^{2})}/{G_{M}^{p}(Q^{2})}\approx 1.35$. As for the isoscalar
nucleon form factor~\cite{CP2}, we find at $Q^{2}\approx 10
GeV^{2}$, $Q^{4}G_{A}^{(s)}(Q^{2})=0.83 GeV^{4}$ for
$\Lambda_{QCD}=100 MeV$ and $Q^{4}G_{A}^{(s)}(Q^{2})=1.34 GeV^{4}$
for $\Lambda_{QCD}=180 MeV$. Assuming isospin invariance, we combine
these results with those for $g_{A}$ to obtain
${G_{A}^{(s)}}/{G_{A}^{(3)}}\approx 1.85$, where $G_{A}^{(3)}$ is
the isovector axial-vector nucleon form factor. If a dipole form
$G_{A}^{(s)}(Q^{2})={G_{A}^{(s)}(0)}/{(1+Q^{2}/M_{AS}^{2})^{2}}$,
with $G_{A}^{(s)}(0)=0.38$ from $SU(6)$, is used to describe the
$Q^{2}$ dependence of $G_{A}^{(s)}$~\cite{CP2}, then, in the
high-$Q^{2}$ region, our model yields $M_{AS}=(1.15-1.22) GeV$
for $\Lambda_{QCD}=100 MeV$ and $M_{AS}=(1.27-1.37) GeV$ for
$\Lambda _{QCD}=180MeV$. These results might be relevant to studies
concerning the strange-quark content of the nucleon.

The last issue we address in this work are the exclusive
decays of the charmonium levels $^3S_{1}$, $^3P_{1}$, and $^3P_{2}$
into $p\bar p$. Such calculations have been carried out by several
authors~\cite{CZ,COZ2,A,BL,DTB} within the QCD convolution
framework. We here follow~\cite{COZ2}. We consider first the
two $\chi_{c}(1P)$ states. The branching ratio for the decay of the
$J^{CP}=1^{++}$ state into $p\bar p$ is given by
\begin{equation}
  BR\Biggl({{^3P_{1}\to p\bar p}\over {^3P_{1}\to all}}\Biggr)
  \approx {{0.75}\over{\ln({\bar M}/{\Delta})}}
  {{16\pi^{2}}\over{729}}{\Bigg \vert {{f_{N}}\over{{\bar M}^{2}}}
 \Bigg \vert}^{4}{M_{1}^{2}},
\end{equation}
where $\bar M\approx 2 m_{c} \approx 3 GeV$ and $\Delta =0.4 GeV$
(the last value from~\cite{BARB}-see also~\cite{NOV}). The
nonperturbative content of Eq. (7) is due to $f_{N}$ and the
amplitude for the process $^3P_{1}\to p \bar p$, $M_{1}$,
which involves $\Phi_{N}$. Using (6) the calculation of $M_{1}$
yields $M_{1}^{het}=99849.6$ and as a result
$BR(^3P_{1}\to p\bar p/^3P_{1}\to all)=0.77\times 10^{-2}\%$, which
is in accordance with the experimental value
$(0.5-1.0)\times 10^{-2}\%$~\cite{PDG}.

The analogous expression to (7) for the $J^{PC}=2^{++}$ state
has the form
\begin{equation}
 BR\Biggl({{^3P_{2}\to p\bar p}\over {^3P_{2}\to all}}\Biggr)
  \approx 0.85 (\pi\alpha_{s})^{4}
  {{16}\over{729}}{\Bigg \vert {{f_{N}}\over{{\bar M}^{2}}}
  \Bigg \vert}^{4}{M_{2}^{2}}\;,
\end{equation}
which is Eq. (20) of~\cite{COZ2} with an obvious minor
correction. For the heterotic NDA
we find $M_{2}=515491.2$. Setting $\alpha_{s}(m_{c})=0.210\pm 0.028$
(see third paper of~\cite{BARB}), we then obtain from (8)
$BR({^3P_{2}\to p\bar p}/{^3P_{2}\to all})=
0.89\times 10^{-2}\%$ in remarkable agreement with
the measured value $(0.90{{+0.41}\atop{-0.26}}\pm0.19)
\times 10^{-2}\%$~\cite{PDG}.

The partial width of the $J^{PC}=1^{--}$ state into $p\bar p$ is
\begin{equation}
  \Gamma(^3S_{1}\to p\bar p) = (\pi\alpha_{s})^{6}{{1280}\over{243\pi}}
  {{{\vert f_{\psi}\vert}^{2}}\over{\bar M}}
    {\Bigg \vert {{f_{N}}\over{{\bar M}^{2}}}
  \Bigg \vert}^{4}{M_{0}^{2}}\;,
\end{equation}
where $f_{\psi}$ determines the value of the $^3S_{1}$-state wave
function at the origin. Its value can be extracted from the leptonic
width $\Gamma(^3S_{1}\to e^{+}e^{-}) = (4.72\pm 0.35)keV$~\cite{PDG}
via the Van Royen-Weisskopf formula. The result is
$\vert f_{\psi}\vert =383 MeV$ with $m_{J/\psi}$ equal
to its experimental value. The heterotic solution leads to
$M_{0}=13726.8$. Then, using the previous parameters, it follows
that
$\Gamma(^3S_{1}\to p\bar p) = 0.12 keV$. From experiment
\cite{PDG} it is known that $\Gamma(p\bar p)/\Gamma_{tot}=
2.16\pm0.11\times10^{-3}$ with $\Gamma_{tot}=(68\pm 10)keV$,
so that $\Gamma(^3S_{1}\to p\bar p) = 0.15 keV$ in
excellent agreement with the model prediction. For the
branching ratio we find
$BR({^3S_{1}\to p\bar p}/{^3S_{1}\to all})=1.76\times 10^{-3}$,
or $1.40\times 10^{-3}$ if the new~\cite{HP} value $\Gamma_{tot}=
85.5{{+6.1}\atop {-5.8}}keV$ is used. We emphasize that the
model predictions for all considered charmonium decays are obtained
with the {\it same} values of the various parameters.

By considering a number of exclusive reactions involving the
NDA $\Phi_{N}$, we have effected
that a novel solution, which we call heterotic, leads to
predictions which are corroborated by experiment. This solution
is consistent with the SR requirements up to the third order
and allows for the possibility to analyze the form-factor data
with the assumption that asymptotically
$\vert G_{M}^{n}\vert /G_{M}^{p}\le 0.1$.
We have pursued our approach, in spite the objections raised by
Isgur and Llewellyn Smith~\cite{ILS} and also by Radyushkin~\cite{RAD}.
We nevertheless believe that our model provides
a useful and predictive tool for phenomenological studies.
\figure{The proton magnetic form factor calculated with the
        heterotic distribution amplitude in comparison with
        the data.\label{Figr2}}
\figure{The neutron magnetic form factor calculated with
           the heterotic distribution amplitude in comparison
           with the data.\label{Figr3}}
\figure{Comparison with available data of the transition form factor
           $\gamma p \Delta^{+}$
           calculated with the heterotic nucleon distribution
           amplitude and two different models for the $\Delta^{+}$
           resonance, as explained in the text.\label{Figr4}}
\narrowtext
\begin{table}
\caption{Moments $n_{1}+n_{2}+n_{3}\le 3$ of the heterotic
         nucleon distribution amplitude $\Phi_{N}$ in
         comparison with the sum-rule constraints}
\begin{tabular}{ccc}
         Moments ${(n_{1}n_{2}n_{3})}$ & Sum rules  &
         $\Phi_{N/het}^{(n_{1}n_{2}n_{3})}$ \\
\tableline
         (000)   & 1                & 1       \\
         (100)   & 0.54---0.62      & 0.572   \\
         (010)   & 0.18---0.20      & 0.184   \\
         (001)   & 0.20---0.25      & 0.244   \\
         (200)   & 0.32---0.42      & 0.338   \\
         (020)   & 0.065---0.088    & 0.066   \\
         (002)   & 0.09---0.12      & 0.170   \\
         (110)   & 0.08---0.10      & 0.139   \\
         (101)   & 0.09---0.11      & 0.096   \\
         (011)   & --0.03---0.03     & --0.021  \\
         (300)   & 0.21---0.25      & 0.21    \\
         (030)   & 0.028---0.04     & 0.039   \\
         (003)   & 0.048---0.056    & 0.139   \\
         (210)   & 0.041---0.049    & 0.079   \\
         (201)   & 0.044---0.055    & 0.049   \\
         (120)   & 0.027---0.037    & 0.050   \\
         (102)   & 0.037---0.043    & 0.037   \\
         (021)   & --0.004---0.007   & --0.023  \\
         (012)   & --0.005---0.008   & --0.007  \\
\end{tabular}
\label{Tab1}
\end{table}
\end{document}